\documentclass[preprint]{aastex631}
\submitjournal{PASP}
\shorttitle{Search Capability for NEOs with the WFST}
\shortauthors{Lu, J.-Q. et al.}

\graphicspath{{./}{figs/}}

\begin{document}

\title{Search Capability for Near-Earth Objects with the Wide Field Survey Telescope}

\author[0009-0008-4604-9674]{Jun-Qiang Lu}
\affiliation{Department of Astronomy, University of Science and Technology of China, Hefei 230026, China}
\affiliation{School of Astronomy and Space Science, University of Science and Technology of China, Hefei 230026, China}

\correspondingauthor{Jun-Qiang Lu; Lu-Lu Fan}
\email{ljq1999@mail.ustc.edu.cn; llfan@ustc.edu.cn}

\author[0000-0003-4200-4432]{Lu-Lu Fan}
\affiliation{Department of Astronomy, University of Science and Technology of China, Hefei 230026, China}
\affiliation{School of Astronomy and Space Science, University of Science and Technology of China, Hefei 230026, China}
\affiliation{Institute of Deep Space Sciences, Deep Space Exploration Laboratory, Hefei 230026, China}

\author[0000-0003-4721-6477]{Min-Xuan Cai}
\affiliation{Department of Astronomy, University of Science and Technology of China, Hefei 230026, China}
\affiliation{School of Astronomy and Space Science, University of Science and Technology of China, Hefei 230026, China}

\author[0009-0007-8133-5249]{Shao-Han Wang}
\affiliation{Department of Astronomy, University of Science and Technology of China, Hefei 230026, China}
\affiliation{School of Astronomy and Space Science, University of Science and Technology of China, Hefei 230026, China}

\author[0000-0003-4122-6949]{Bing-Xue Fu}
\affiliation{Department of Astronomy, University of Science and Technology of China, Hefei 230026, China}
\affiliation{School of Astronomy and Space Science, University of Science and Technology of China, Hefei 230026, China}

\author[0000-0002-7660-2273]{Xu Kong}
\affiliation{Department of Astronomy, University of Science and Technology of China, Hefei 230026, China}
\affiliation{School of Astronomy and Space Science, University of Science and Technology of China, Hefei 230026, China}
\affiliation{Institute of Deep Space Sciences, Deep Space Exploration Laboratory, Hefei 230026, China}

\author[0000-0003-0694-8946]{Qing-Feng Zhu}
\affiliation{Department of Astronomy, University of Science and Technology of China, Hefei 230026, China}
\affiliation{School of Astronomy and Space Science, University of Science and Technology of China, Hefei 230026, China}

\begin{abstract}

Wide Field Survey Telescope (WFST), with a powerful sky survey capability in the northern hemisphere, will play an important role in asteroid searching and monitoring. 
However, WFST is not a telescope dedicated to near-Earth asteroids (NEOs) searching. 
In order to improve the efficiency of finding NEOs on the premise of meeting the needs of other scientific research, we ran mock observations for WFST to study its search capability for NEOs.
The NEO population model, the WFST detection model and site conditions are taken into account in our simulations.
Based on the original scheduling scheme, we present two new schemes.
Compared to the original scheme, the optimized scheme can improve the search capability of known and unknown NEOs by 100\% and 50\%.
We also emphasized the importance of trailing loss and proposed an improved effective field of view model.
In addition, it is predicted that adopting the clear-day ratio of 0.7 and the optimized scheme, during one year of regular survey, for NEOs with absolute magnitude from 17 to 25, WFST can provide tracklets for about 1800 NEOs if their orbits are known, and in the case of blind search, more than 600 NEOs can be found by WFST.
The new schemes provide valuable reference and suggestions for the WFST's regular survey strategy.

\end{abstract}


\section{Introduction}\label{Intro}

Near-Earth Objects (NEOs) are usually defined as asteroids or comets with a perihelion distance ($q$) less than 1.3 AU. Although rare, the impact of NEOs can pose a great threat to Earth's civilization \citep{chapman_hazard_2004}. Studies show that the end-Cretaceous extinction was related to the impact of near-earth objects \citep{alvarez_extraterrestrial_1980,nicholson_3d_2024}. The Chelyabinsk incident in 2013 \citep{brown_500-kiloton_2013,popova_chelyabinsk_2013} caused widespread concern about near-earth asteroids. However, the first step of planetary defense is the discovery and monitoring of near-earth asteroids.

In recent decades, a large number of NEOs have been discovered and cataloged at the Minor Planet Center (MPC) by survey projects such as the Catalina Sky Survey \citep{christensen_catalina_2019}, Pan-STARRS \citep{wainscoat_pan-starrs_2016}, ATLAS \citep{heinze_neo_2021}, NEOWISE \citep{mainzer_population_2014,masiero_characterization_2021}, and ZTF \citep{bellm_zwicky_2019}. In the near future, the Vera Rubin Observatory \citep[LSST,][]{jones_large_2018} and NEO Surveyor \citep{mainzer_near-earth_2023} will also have a great impact on the search and monitoring of NEOs.

The Wide Field Survey Telescope (WFST) is an optical telescope jointly built by the University of Science and Technology of China  (USTC) and the Purple Mountain Observatory (PMO), Chinese Academy of Sciences (CAS), located at an altitude of 4200 m on the Saishiteng Mountain near Lenghu Town, Qinghai Province, China. It has a primary mirror diameter of 2.5 m, an active optics system, six filters ($u$,$g$,$r$,$i$,$z$,$w$) and a mosaic camera consisting of nine 9K $\times$ 9K CCDs yielding a 3-degree diameter field of view.\footnote{\url{https://wfst.ustc.edu.cn/}}

It is expected that WFST will be an effective tool to detect supernovae, Tidal Disruption Events (TDEs) and some other time-domain sciences \citep[e.g.,][]{hu_prospects_2022,lin_prospects_2022,wang_study_2022,liu_target--opportunity_2023}. Some tools, methods, and pipelines for WFST have been developed \citep[e.g.,][]{xu_new_2022,liang_kilonova-targeting_2023,liu_classification_2025} or are being developed. As a survey telescope, solar system science, including monitoring and searching for NEOs, is naturally one of the scientific goals of WFST \citep{wang_sciences_2023}. However, WFST is not a telescope dedicated to NEOs, and its regular sky survey usually will not visit the same field three or more times in a single night. Under such a cadence, how to optimize the survey strategy to discover and monitor more NEOs has become a problem. In this paper, we have modeled the NEO population, the Lenghu site, and the ability of WFST, evaluated and compared the detection capability of WFST to NEOs under several different survey strategies through mock observations, which provided some reference guidance for formulating the observation plan of WFST's upcoming regular survey.

The structure of this paper is as follows. In Section \ref{methods}, we introduce the models and methods used in our simulations. The basic requirements of the survey strategy for WFST and several specific strategies we tried when running the simulations are elaborated in Section \ref{strategy}. Then, we displayed the results of our mock observations and some related discussions in Section \ref{results}. Finally, we summarize our findings in Section \ref{summary}.

\section{Methods}\label{methods}

\subsection{NEOs Population}

Some four-dimension models \citep{bottke_understanding_2000,bottke_debiased_2002,granvik_super-catastrophic_2016,granvik_debiased_2018} have described the debiased orbits distribution and absolute magnitude distribution of NEOs, in the form of a function of semi-major axis ($a$), eccentricity ($e$), inclination ($i$) and absolute magnitude ($H$). Recently, \cite{nesvorny_neomod_2023,nesvorny_neomod_2024,nesvorny_neomod_2024-1,deienno_debiased_2025} also presented updated NEO models after this work began. Although the orbital angular elements (longitude of the ascending node $\Omega$, argument of perihelion $\omega$, mean anomaly $M_0$) of NEOs are not statistically uniformly distributed \citep{jeongahn_non-uniform_2014}, all these studies think that they are negligible for their purposes.

In this paper, we use Granvik's NEOs model \citep{granvik_debiased_2018}, which contains 802,000 simulated objects with $q<1.3\,\rm{AU}$, $a<4.2\,\rm{AU}$ and $17 < H < 25$, corresponding to 1 km $\sim$ 25 m diameter assuming the albedo is 0.25. The total number of 802,000 roughly equals the actual number of NEOs given by their unbiased H-frequency distribution (HFD) within this range. We also divide them into Amors, Apollos, Atens, and Atiras according to their orbits.\footnote{\url{https://cneos.jpl.nasa.gov/about/neo_groups.html}}
Furthermore, we calculated their Earth Minimum Orbit Intersection Distance (MOID) using \citet{wisniowski_fast_2013}'s method, and defined NEOs with MOID $\leq$ 0.05 AU and $H \leq 22$ as Potentially Hazardous Asteroids (PHAs).

\subsection{Orbital Propagation}

We regard the orbital elements provided by Granvik's model as the orbital elements at a selected initial moment, and then propagate the orbits of these 802,000 objects to another one-year period that we selected, and generate an ephemeris every night. In this step, we assume that all objects obey the simple Kepler motion law, which is reasonable because we don't care about the strict orbit of any virtual object, we only care about their statistical position distribution.

After getting the ephemeris of every night, we use the Cartesian coordinates in the ephemeris to get the celestial coordinates and apparent motion speed of each virtual object, and then we assume that every object moves approximately linearly on the celestial sphere at a constant speed this night to get the position of the object in the middle of each exposure. 
This step is the most time-consuming step in the whole simulation because it requires a lot of calculation, reading, and writing. If an accurate ephemeris is generated for these 802,000 objects at every moment, it will make the simulation too time-consuming, and it is not necessary to do so. Therefore, we choose to use the linear motion approximation in a single night, which is reasonable because many algorithms search for asteroids by connecting the sources that show linear motion in a single night. Even so, it is estimated that it will take about 20 hours to run each simulation, so we shortened the time of running each simulation to about 2 hours by parallelizing the program.

\subsection{Asteroid Detection}

\subsubsection{Magnitudes Calculation}
The $5\sigma$ limiting magnitudes of WFST in several different moon phases are given by \citet{lei_limiting_2023} when airmass=1.2, in 30 s exposure. 
We used cubic splines to interpolate them into continuous moon phases and then corrected for atmospheric extinction for different altitude angles.
Atmospheric extinction can be estimated by the formula $\Delta m=k_{ext}\cdot X$.
In reality, the atmospheric extinction coefficient $k_{ext}$ is variable.
For simplicity, we used the median value of each band of WFST during photometric nights (Cai, M.-X., et al., in preparation).
The airmass $X$ can be calculated using the formula $X=(\cos(z) +0.025e^{-11\cos (z)})^{-1}$ by \citet{rozenberg_twilight_1966}, where $z$ is the zenith distance.
For the 90 s exposure, because there is only data with airmass = 1.2 and moon phase = 0 \citep{wang_sciences_2023}, we simply add a constant correction term to the 30-s exposure limiting magnitude for each band to calculate their limiting magnitudes.

The apparent magnitude of the asteroids in Johnson's $V$ band can be model as
\begin{equation}
    V=H+\delta+5\log_{10}(r\Delta)-2.5\log_{10}\phi(\alpha),
\end{equation}
where $H$ is the absolute magnitude, $\delta$ is the magnitude variation caused by rotation, $r$ is the heliocentric distance (in AU), $\Delta$ is the geocentric distance (in AU), and $\phi(\alpha)$ is the phase function, which varies with the Sun–asteroid–Earth angle $\alpha$.
In this paper, we did not consider $\delta$, because the randomness caused by this term can be represented by the detection fading effect in Sec \ref{detection} to some extent.
The two-parameter $H$-$G$ magnitude system developed by \citet{bowell_application_1989} is used to calculate $\phi(\alpha)$, assuming $G$ = 0.15.

If we know the reflectance spectra of an asteroid, we can convert the apparent magnitude of the asteroid in the V-band into the apparent magnitude of other bands.
According to the fractional abundances and albedos of different taxonomic complexes for NEOs \citep{stuart_bias-corrected_2004},
we divide the near-earth asteroids into the C-like group (61.11\%) with low albedo and the S-like group (38.19\%) with high albedo depending on whether the albedo is less than 0.2.
Then we assume that WFST and LSST have the same transmission curve in the $u$, $g$, $r$, $i$, and $z$ bands, and we use the color transformation of LSST \citep{jones_large_2018} to calculate the apparent magnitude of the asteroids in each band. There is no need to consider the color conversion of the $w$-band because the $w$-band will not be used for WFST's regular survey.

Due to the high proper motion speed of NEOs, it is necessary to consider the trailing losses. It can be described \citep{jones_large_2018} by the function
\begin{eqnarray}
    \Delta m & = & 1.25\log_{10} (1+cx^2),\\
    x & = & \frac{vt}{24\theta},
\end{eqnarray}
where $c=0.42$ is derived for LSST, $v$ is the velocity (in deg/day), $t$ is the exposure time (in seconds) and $\theta$ is the FWHM (in arcseconds). For WFST, the value of $c$ is not determined, so we simply use the same value as LSST. Since no other reference values are available, this choice might introduce some error. However, as discussed in Section \ref{results}, considering trailing losses is necessary to avoid overestimating the number of NEOs detected in this study. The median seeing of 0.75 arcseconds at the Lenghu site \citep{deng_lenghu_2021} is used as the value of $\theta$.

\subsubsection{Source Detection}\label{detection}

The field of view (FoV) of WFST is a circle with a diameter of 3 degrees, and nine CCDs are arranged in a $3\times3$ array. Because the side length of the CCD array is smaller than the diameter of the circle and there are gaps between the CCDs, only part of the circle will appear in the final image, which is called effective FoV (eFoV). The ratio of eFoV to FoV is called the filling factor. Since the widths of the gaps between the CCD arrays are not all the same, we used a Monte Carlo simulation of a series of real exposure images and found the filling factor value to be 0.842.

An easy way to determine whether a simulated object is in the eFoV is to generate a random number uniformly distributed between 0 and 1 when the object falls in the FoV circle, and accept the object if it is less than 0.842, otherwise reject it. But there is an underlying problem with this ``random eFoV model". If a telescope has a small filling factor, the probability that an asteroid is observed many times during the same night might be underestimated, because it is not easy to run too far from their previous positions in just a few hours.

\begin{figure*}[htb!]
  \centering
   \includegraphics[width=.9\textwidth]{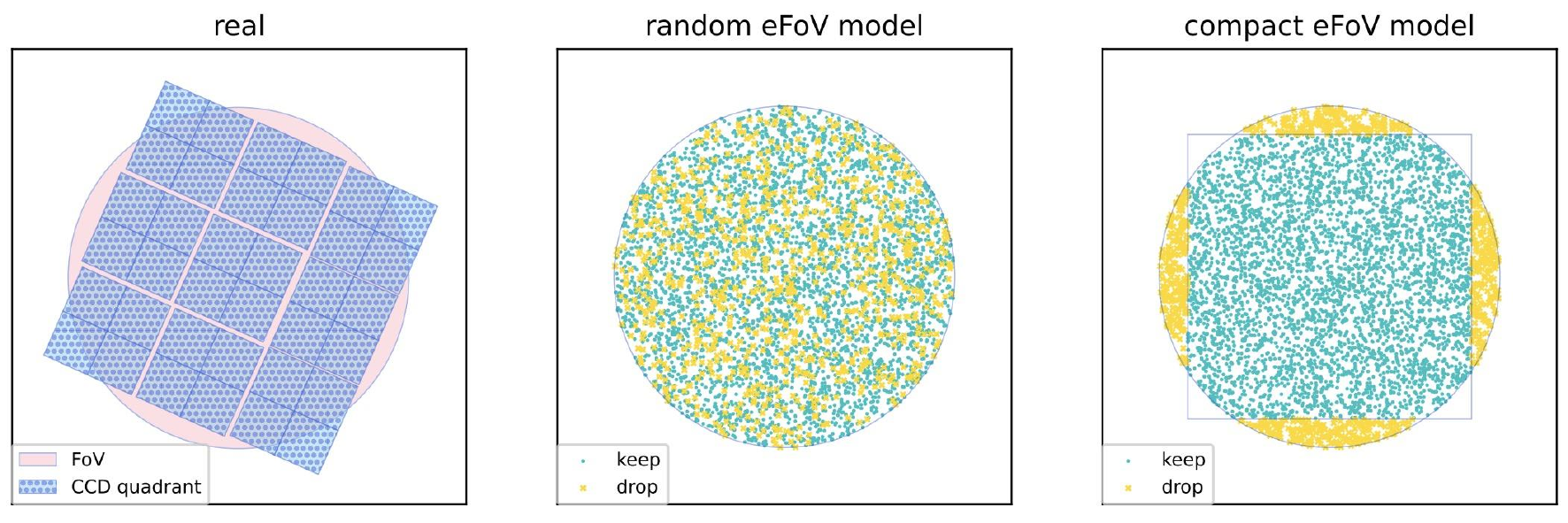}
    \caption{Three eFoV models of WFST. On the left is the arrangement of WFST's CCDs and FoV, and each CCD is divided into four quadrants in data processing. In actual observations, the direction of CCDs is hardly parallel to the right ascension and declination. Only the sources that fall on the FoV and CCD at the same time can be effectively detected. In the middle and on the right are random eFoV model and compact eFoV model respectively. Cyan dots will be retained, while golden dots will be removed.}
    \label{fig:eFoV_models}
\end{figure*}

In order to improve this, we assume that nine CCDs are arranged closely without gaps, and adjust the side length of the CCDs array to meet the filling factor of about 0.842. In this ``compact eFoV model", whether an asteroid falls in the eFoV every time is no longer random, but determined by its position. Although ignoring the gaps may slightly overestimate the probability of an asteroid forming tracklets, we think it is a balance between using the real but cumbersome eFoV and using the random eFoV. All these eFoV models are illustrated in Figure \ref{fig:eFoV_models}.

\begin{figure}[htb!]
  \centering
   \includegraphics[width=.8\linewidth]{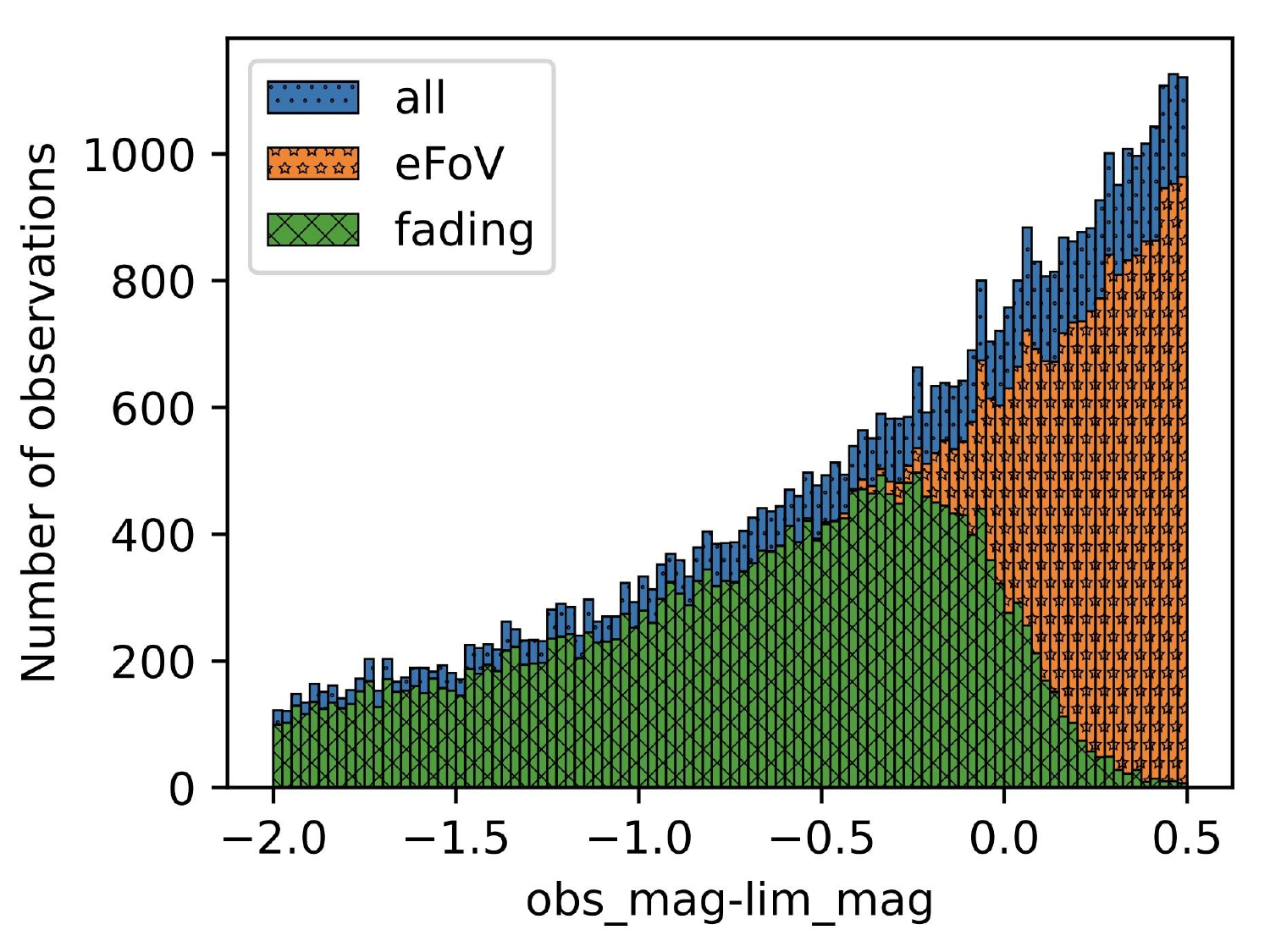}
    \caption{Depiction of our detection model. At first, all the observations that are 0.5 mag larger than the limiting magnitude and fall within the radius of 1.5 degrees of FoV are preserved (blue). Then, according to our eFoV model, about 84\% of the sources are preserved (orange). At last, the observations selected by the detection fading function are considered to be detected (green).}
    \label{fig:eFoV_fading}
\end{figure}

As for detection fading, we use a more realistic probability function to judge whether a source is detected, instead of directly cutting off at the limiting magnitude. Like \citet{jones_large_2018}, we generate a random number R drawn from a uniform distribution between 0 and 1 for each source, and a source is accepted if
\begin{equation}
    R<(1+\exp \frac{m_{obs}-m_{lim}}{\sigma})^{-1}
\end{equation}
where $m_{obs}$ is the observed magnitude, $m_{lim}$ is the limiting magnitude, and $\sigma=0.1$ is a fixed fading factor.Because the limiting magnitude for different signal-to-noise ratios (SNR) is not provided by \citet{lei_limiting_2023}, and the source catalogs generated by the WFST data processing pipeline \citep{cai_25-meter_2025} also uses SNR = 5 as the threshold, all results discussed in this paper are based on an SNR threshold of 5. A depiction of our detection model is shown in Figure \ref{fig:eFoV_fading}.

The night weather in the Lenghu site is divided into two situations: clear and not clear, and we assume that the data in the latter situation is completely unavailable. For convenience in comparing different simulation schemes, in this paper, unless otherwise specified, we consider the case of a 100\% clear-day ratio. The influence of weather conditions on the number of discovered NEOs is also discussed in Section \ref{results}.

\subsubsection{Searching Algorithm}

Traditional asteroids searching methods used in the sky survey, such as Pan-STARRS's Moving Objects Processing System \citep[MOPS,][]{denneau_pan-starrs_2013}, need to visit the same field at least three times in one night. This is because there are many false detections, and it is necessary to link the sources by checking whether the object moves approximately along a straight line. This step is called building ``tracklets". In recent years, some new methods have been proposed, such as HelioLinC \citep{holman_heliolinc_2018}, ZMODE \citep{masci_zwicky_2019}, THOR \citep{moeyens_thor_2021}, CANFind \citep{fasbender_exploring_2021}, and Fink-FAT \citep{montagner_enabling_2023}.

In most cases, WFST will not visit the same field three times or more in a single night, so a Solar System Objects (SSOs) searching pipeline based on the HelioLinC algorithm has been developed \citep{wang_heliocentric-orbiting_2025}. It can not only find unknown SSOs that have been observed three times or more in a single night, but also find unknown SSOs that have been observed on at least three nights within 14 days, with a minimum of two visits per night. For known asteroids, they will be matched according to the ephemeris, and two appearances in a single night are enough to meet the requirements for submission to the MPC\footnote{\url{https://www.minorplanetcenter.net/iau/info/Astrometry.html}}. In this study, we consider an NEO found if it meets the criteria set by the searching pipeline for finding unknown SSOs. It should be noted that while this is sufficient for submission to the MPC, it may not be sufficient for obtaining a provisional designation for the asteroid, as additional observations are typically required for confirmation.

\section{Survey Strategy}\label{strategy}

\subsection{Survey Proposals}

There are two key survey proposals for WFST's 6-year regular survey starting at the end of 2024,
the Wide-Field Survey (WFS) and the Deep High-cadence $u$-band Survey (DHS), each accounting for 45\% of the observation time.
Their basic requirements have been elaborated by \citet{wang_sciences_2023}, and we summarize them in Table \ref{tab:proposals}.

\begin{deluxetable*}{cccCC}
\tablecaption{Basic requirements for different survey proposals of WFST\label{tab:proposals}}
\tablewidth{0pt}
\tablehead{\colhead{Proposal} & \colhead{Time occupied} & \colhead{Exposure time} & \colhead{Area} & \colhead{Filters}}
\startdata
WFS & 45\% & 30 s & \sim 8000\,\rm{deg^2} & u/g/r/i (FLI < 0.65),\,g/r/i (FLI\geq 0.65)\\
DHS & 45\% & 90 s & \sim 720\,\rm{deg^2} & u+g/r/i/z (FLI < 0.5),\,g/r/i/z (FLI\geq 0.5)
\enddata
\tablecomments{The optional bands of WFS are determined by whether it is bright night or not. In this work, dark nights, gray nights, and bright nights are divided by Fractional Lunar Illumination ($FLI$) defined by \citet{skillen_new_2002}.
When $0\leq FLI < 0.25$, it is dark.
When $0.25\leq FLI < 0.65$, it is a gray night.
When $0.65\leq FLI < 1$, it is a bright night.
The DHS bands are determined by whether it is within 7 days before and after a new moon. Here we use whether $FLI$ is less than 0.5 as an alternative criterion.
}
\end{deluxetable*}

\begin{figure*}[htb!]
  \centering
   \includegraphics[width=.9\textwidth]{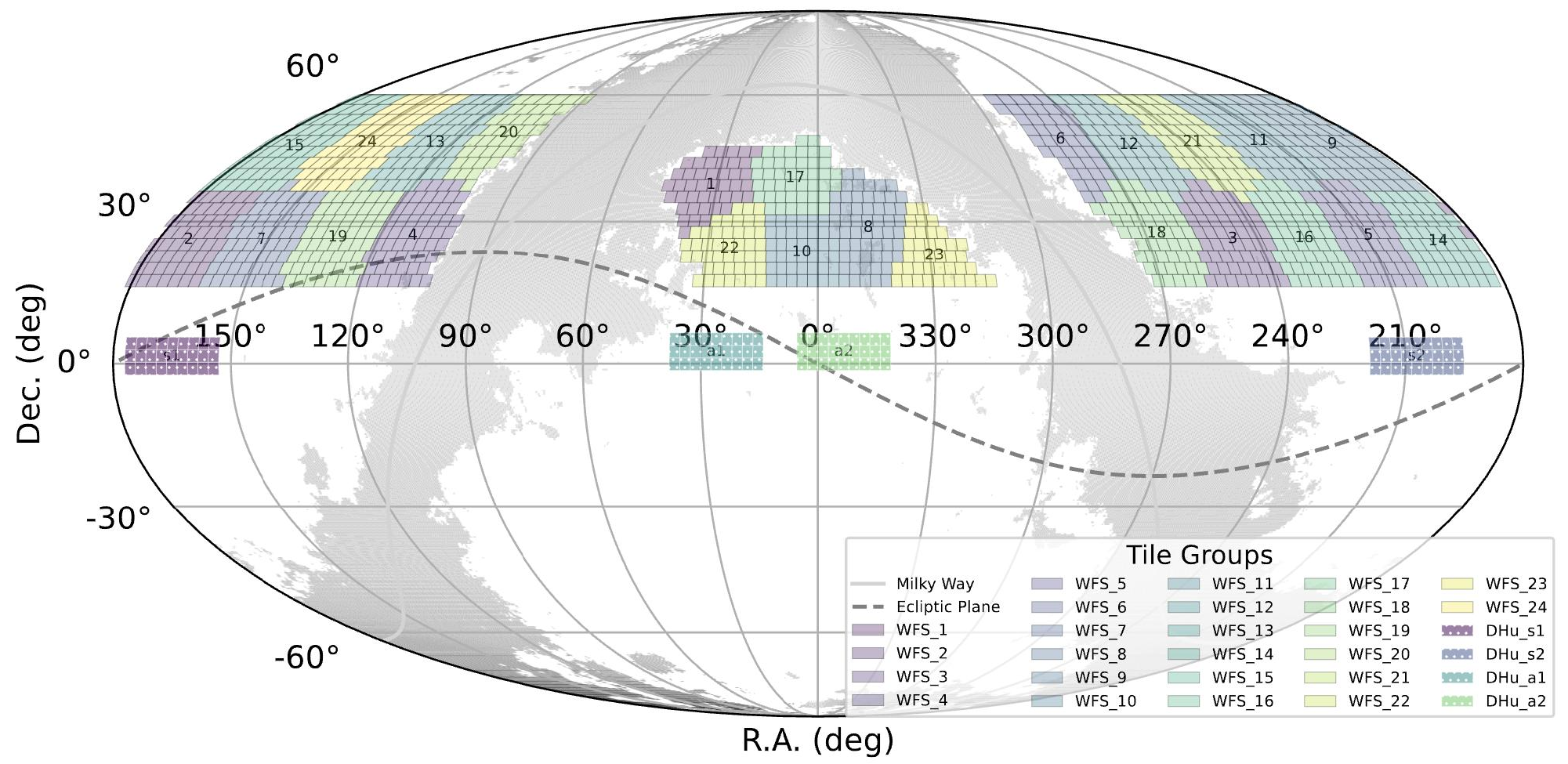}
    \caption{Survey regions of WFST. The light gray area is the area where the extinction of the Milky Way is significant, and the dashed line is the position of the ecliptic plane. The 4 DHS tile groups are all near the celestial equator, which are represented by dark colors. The 24 WFS tile groups are all located in the area where the declination is between 15 and 60 degrees.}
    \label{fig:survey_regions}
\end{figure*}

The sky region with a declination between 15 degrees and 60 degrees and $E(B-V)<0.15$ is selected for WFS, where $E(B-V)$ is the galactic reddening \citep{schlegel_maps_1998} used to exclude the sky region with significant Milky Way extinction. 
A total of 1,086 tiles are divided into 24 tile groups \citep{chen_basic_2023} using the K-Means algorithm, and each of the tile groups contains about 50 tiles.
Each tile covers an area of about 6 square degrees and its center coincides with a WFST pointing.
In addition, DHS will regularly monitor four sky regions near the celestial equator, and each of the sky regions consists of 27 tiles.
All these tile groups are shown in Figure \ref{fig:survey_regions}.

\subsection{Scheduling}\label{sec:Scheduling}

A basic scheduling strategy is given by \citet{chen_basic_2023}.
This scheduling strategy uses a greedy algorithm to select the best tile group at each moment through five metrics. Each metric is linearly scaled to a score between 0 and 10, and finally the tile group with the highest total score is selected. The five metrics are listed as follows:

\begin{description}

\item [Sky background] Scattered moonlight is the most important factor affecting the optical night-sky background. \citet{noll_atmospheric_2012} extended the scattered moonlight of the V-band by \citet{krisciunas_model_1991} to a spectroscopic version. The effective wavelength of each band of WFST \citep{lei_limiting_2023} is used to calculate the sky background under varying moon conditions at the center of each tile group. Additionally, positions where the angular distance from the Moon is less than 45 degrees will receive a very low score to avoid being observed.

\item [Altitude angle] The altitude angle is related to the airmass and will affect the limiting magnitude of the telescope due to atmospheric extinction, so larger altitude angles are preferred when scheduling observation. Additionally, we will give a very low negative score to tile groups whose altitude angle is less than 30 degrees to avoid being selected.

\item [Angular distance] The angular distance from the current tile group is used to evaluate the slew time of the telescope.

\item [Historical observations] The purpose of this metric is to ensure that the total number of observations of each tile group is relatively uniform.

\item [Recently repeated observations] This metric is used to control the telescope's revisiting cadence of the same tile group that night.
\end{description}  

For the following different runs, we will generate the observation plan by adjusting the scoring rules of the metrics above. The advantage of this is that by making tiny changes to the code, the expected cadence can be met as much as possible without violating the limits of the observing conditions.

\subsection{Generating Observation Plans}

Astronomical night is defined as the period when the Sun is 18 degrees below the horizon. For WFST, we assume that a time interval of 20 seconds is needed between two exposures for readout, slewing, etc. Then we calculated the time required to observe each tile group and found that it was within 50 minutes, so we selected a tile group for observation every 50 minutes using the algorithm mentioned in Sec \ref{sec:Scheduling}. If observation of a tile group is completed in advance, we assume that the remaining time of the tile group is unavailable.

Both WFS and DHS have been taken into account. In the process of allocating tile groups every 50 minutes, there is also a probability of 0.1 that the period is being allocated to ToO (Targets of Opportunity), and we assume that these observation times are not used to search for asteroids. Moreover, to meet the requirement of occupying observation time in Table \ref{tab:proposals}, DHS must have priority to reach 45\% of the total observation time, because there are sometimes no DHS tile groups available for observation due to the limitation of altitude angle. When selecting bands, we will not switch the band unless the current band cannot meet the requirements in Table \ref{tab:proposals} or the tile group has been observed with this band on that day. When switching bands, we will randomly select one from the optional bands with equal probability. The only exception is that the $u$ band of DHS will have a probability of 0.5, and the remaining $g$, $r$, $i$ and $z$ bands will share the remaining probability of 0.5 equally.

Finally, we need to expand the tile groups of each day into a specific observation list, 
in which each line gives the celestial coordinates and the corresponding horizontal coordinates at which time the telescope should point to.

\subsection{Simulation Schemes}

\begin{figure*}[thb!]
    \gridline{
        \fig{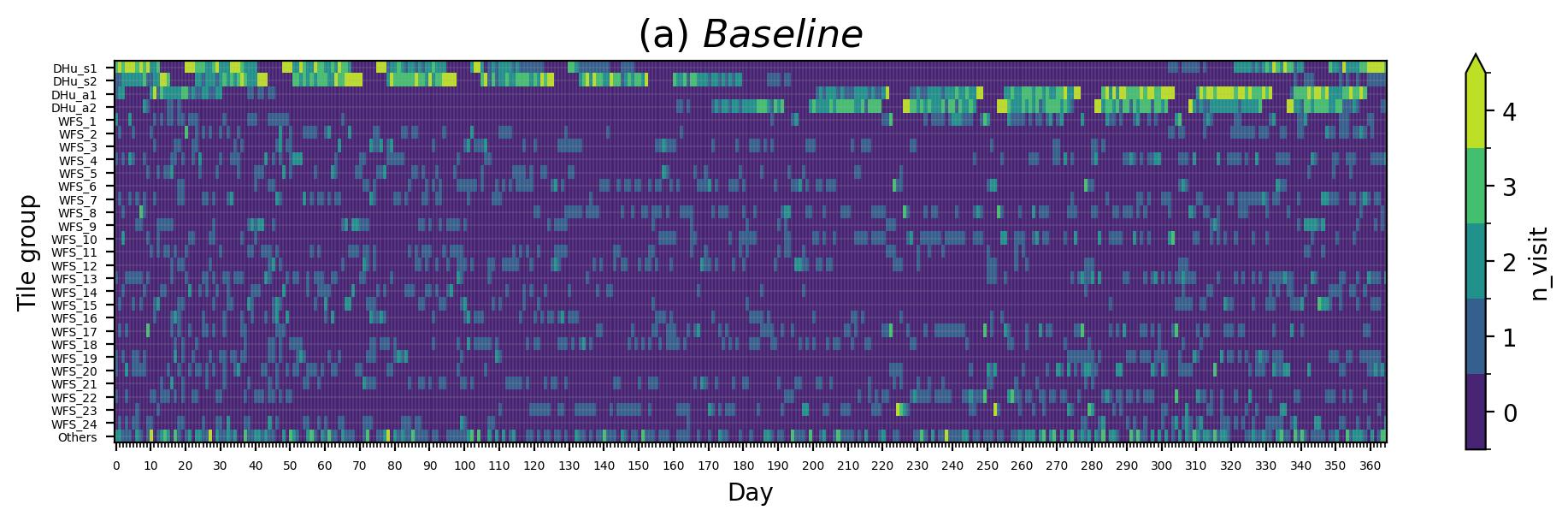}{.9\textwidth}{}
    }\vspace{-3em}
    \gridline{
        \fig{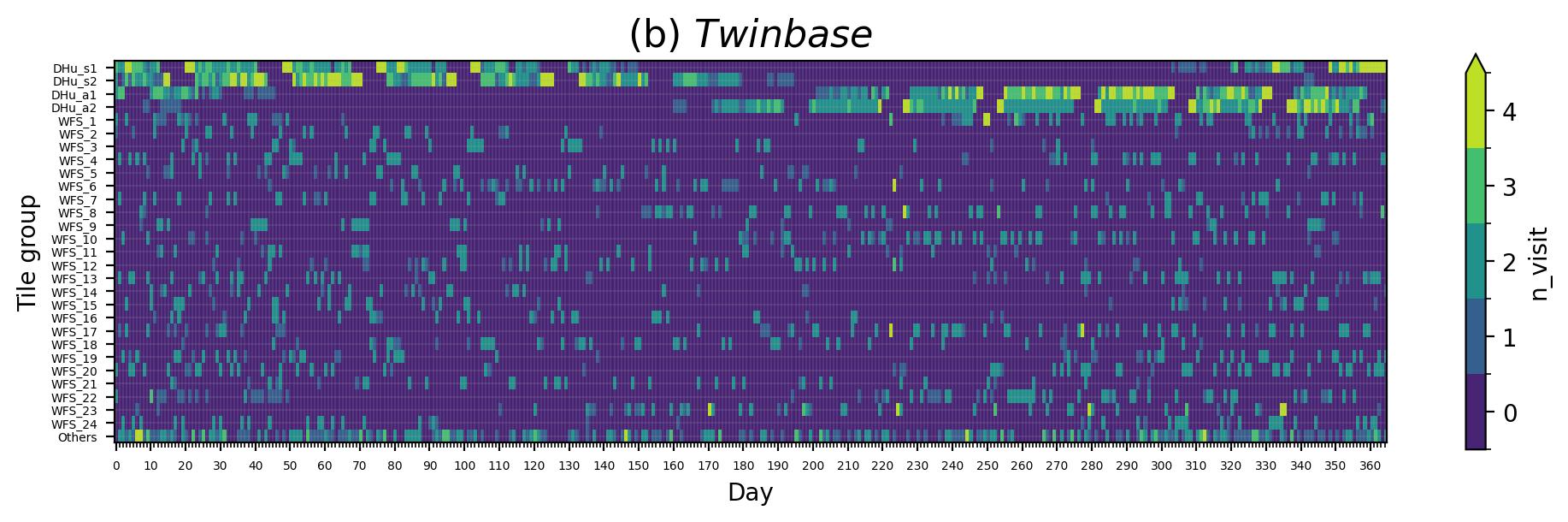}{.9\textwidth}{}
    }\vspace{-3em}
    \gridline{
        \fig{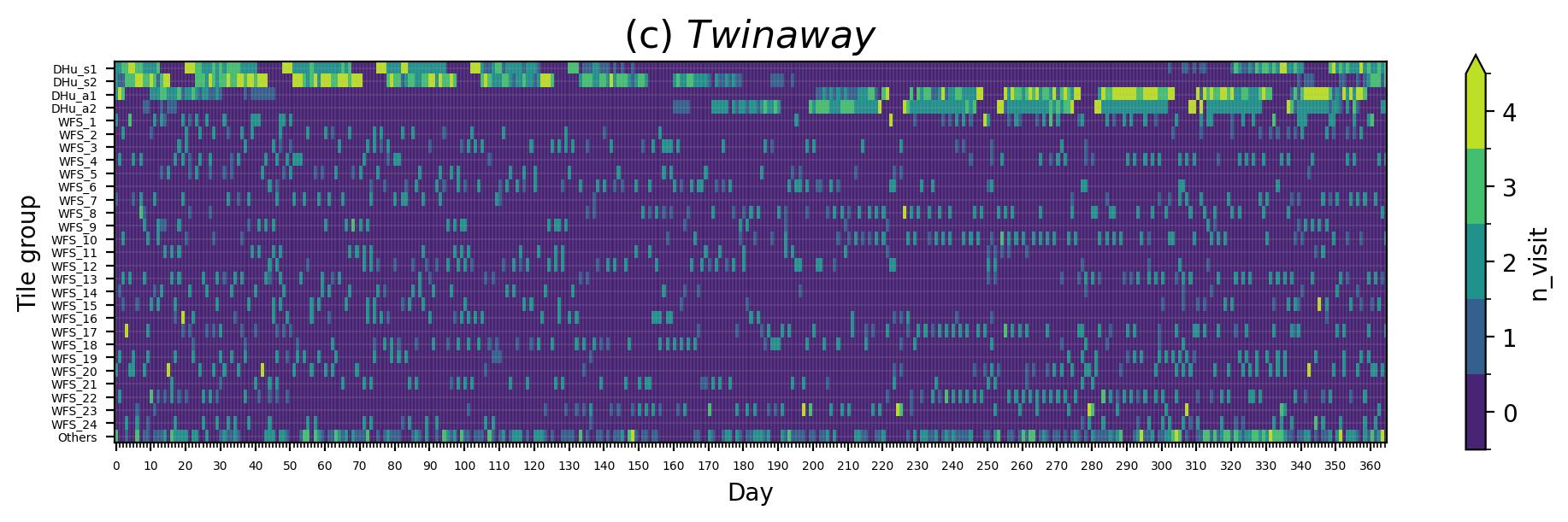}{.9\textwidth}{}
    }\vspace{-3em}
    \caption{The cadence of every tile group in a year. On each panel, the horizontal axis is time and the vertical axis is different tile groups. The top four rows belong to DHS, and the other rows except the last row belong to WFS. The last row is for observing targets of opportunity, which is not considered in this paper. (a) \textit{Baseline}, (b) \textit{Twinbase} and (c) \textit{Twinaway}.}
    \label{fig:cadence}
\end{figure*}

In this paper, mock observations for one year of the three schemes have been tested 10 times. Each scheme aims to optimize the number of discoverable asteroids by adjusting the survey cadence on the basis of the former. The cadence of every tile group in each scheme is shown in Figure \ref{fig:cadence}.

\begin{description}

\item [\textit{Baseline}] This scheme uses the same strategy as \citet{chen_basic_2023}, but adds DHS. When choosing a tile group, the last two tile groups observed that night will be avoided as much as possible by setting their ``recent repeated observations" metric to 0 and setting other tile groups to 10.

\item [\textit{Twinbase}] After noticing that many WFS tile groups in the baseline were only observed once a night, we modified the rules of the metric of ``recent repeated observations". At the beginning of a night, the value of this metric for all tile groups is 5. When a tile group has been observed once, it is set to 10. And if a tile group has been observed twice or more, it is set to 0. This is to make tile groups have paired visits as much as possible, which is very important for discovering NEOs.

\item [\textit{Twinaway}] We found that some tile groups were observed every day for several days, which may not be conducive to obtaining a larger search area. Therefore, in order to let the telescope stay away from the tile groups observed the previous day and cover more sky areas in a few days, we set the ``Historical observations" metric of all the tile groups observed the previous day to 0. And other settings are the same as \textit{Twinbase}.

\end{description}

\section{Results and Discussion}\label{results}

\begin{figure*}[thb!]
    \gridline{\fig{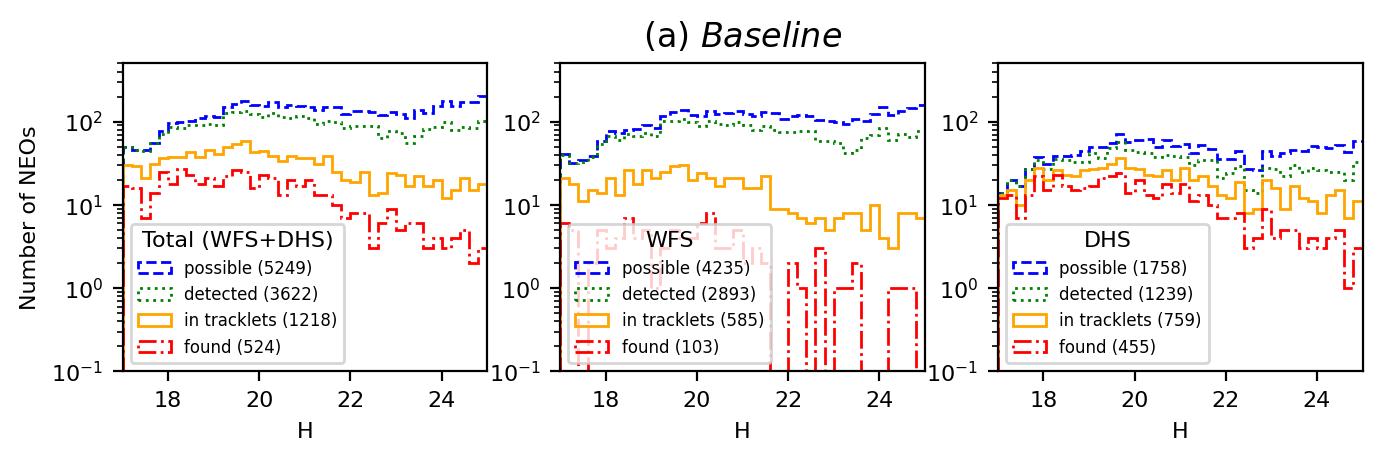}{.9\textwidth}{}}\vspace{-3em}
    \gridline{\fig{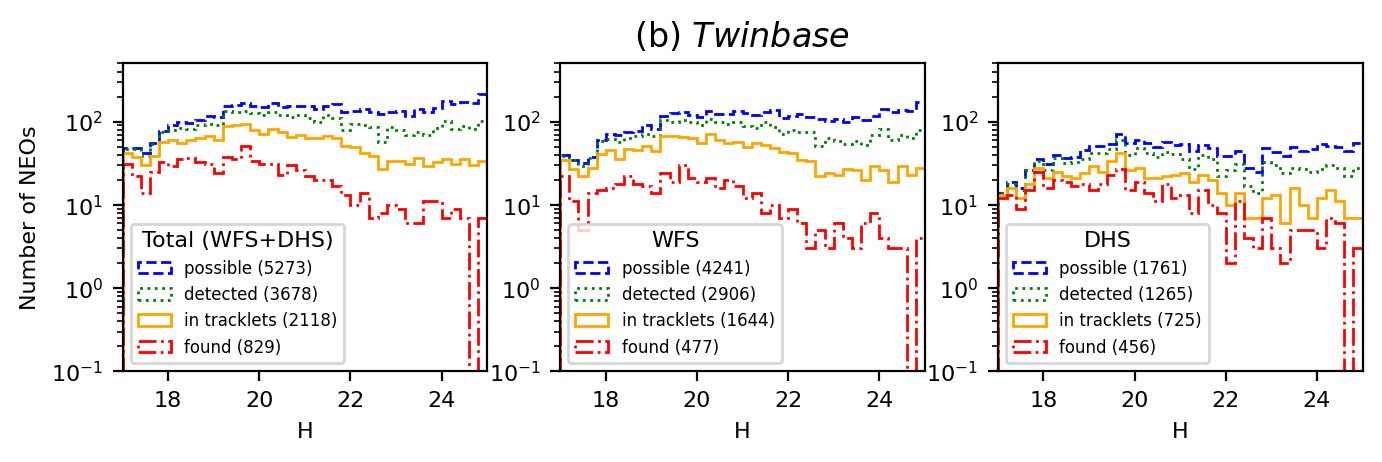}{.9\textwidth}{}}\vspace{-3em}
    \gridline{\fig{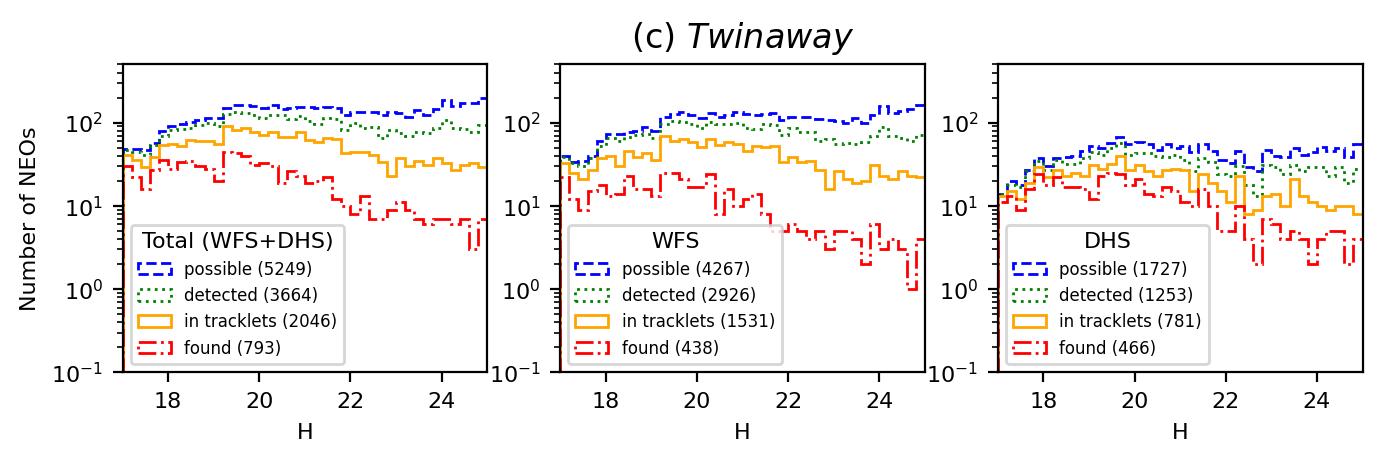}{.9\textwidth}{}}\vspace{-3em}
    \caption{For a one-year simulation, the H frequency distribution for different simulation schemes is shown: (a) \textit{Baseline}, (b) \textit{Twinbase}, and (c) \textit{Twinaway}. Each bin is 0.2 magnitudes wide. The cumulative total number of qualified NEOs for each condition within each plot is indicated in brackets within the legend. The blue, green, yellow and red lines from top to bottom all include only asteroids that fall in the eFoVs. The blue (dashed) lines represent all asteroids that are 0.5 magnitude larger than the limiting magnitude and possible to be detected. The green (dotted) lines represent the asteroids that are thought to have been detected once or more in the simulation. The Yellow (solid) lines represent all asteroids that have been detected at least twice in at least one night (the minimum requirement for forming a tracklet). The red (dashdot) lines represent the asteroids that can theoretically be found by the searching algorithm, having a tracklet with at least 3 detections or at least 3 tracklets in 14 days.}
    \label{fig:HFD}
\end{figure*}

In each subgraph of Figure \ref{fig:HFD}, it can be seen that with the increase of $H$, the width between all two lines increases.
For the top two lines, this means that as the size of asteroids decreases, asteroids are more likely to be fainter and more difficult to detect.
For the middle two lines, the fainter the asteroid, the more difficult it is to observe it twice or more at night, usually in different bands.
Finally, the change in width between the two bottom lines shows that the tracklets of small asteroids are more difficult to be linked, which may be because they are more difficult to detect many times and the observation window is shorter.
All in all, as we can expect, the probability that an NEO can be found by the algorithm decreases with the decrease of its size.

Compared with WFS, the number of NEOs observed by DHS is smaller because of its smaller sky coverage. However, because of the high cadence of revisiting for DHS, more asteroids are able to form tracklets, which are linked into tracks and finally found by the algorithm. In addition, we can also find that DHS is less sensitive to different simulation schemes than WFS.

As we can see in Figure \ref{fig:HFD}, the interval between the green line and the yellow line of WFS in \textit{Baseline} is wide, indicating that a large number of asteroids have isolated detection and cannot form tracklets. As evidenced by Figure \ref{fig:cadence}, most WFS in \textit{Baseline} are only visited once a night when observed (light blue), while most WFS of \textit{Twinbase} and \textit{Twinaway} are visited twice a night (green). Thus, by making each tile group have paired visits every night as much as possible, we can see that the number of asteroids finally found by WFS in Figure \ref{fig:HFD} has increased significantly, and consequently increased the total number of NEOs found. 

However, the change from \textit{Twinbase} to \textit{Twinaway} did not improve the number of NEOs with tracklets or the number of NEOs found significantly. Although the tile group of the previous day was avoided, the tile groups of WFS that are observed within a few days are still limited.
On the one hand, some tile groups may have more opportunities to have three nights of visits in 14 days. On the other hand, because some small-sized NEOs may only have a few days' observation window, three consecutive nights are more conducive to the discovery of such NEOs. The combination of these two aspects leads to \textit{Twinaway}'s no obvious improvement compared with \textit{Twinbase}.

Some factors in the simulation steps will affect the final result. Here, we want to discuss the influence of trailing loss, eFoV models, and weather. In order to evaluate the influence of these factors on the search capability of NEOs, we set 10 different initial epochs to generate ephemeris of the simulated NEOs population and regenerate the observation plans of three schemes for mock observations. 

We tried three kinds of speed limitations, unlimited, removing those whose proper motion is faster than 2 degree/day, and restricting the sources that can be detected by trailing loss. In the first two cases, the number of NEOs that can be found in the end is several times greater than that in the last one. This is because the number of NEOs is roughly distributed with the power of size, and as the size of NEOs decreases, we find that NEOs brighter than the limiting magnitude have higher proper motion speed. Looking at the HFD of NEOs finally found (corresponding to the red line in Figure \ref{fig:HFD}) by the first two kinds of limitations, they did not increase first and then decrease as in Figure \ref{fig:HFD}, but continued to increase in the whole range $H$ from 17 to 25, which did not meet our expectations and experience. Generally speaking, for NEOs, especially for small NEOs, the magnitude loss caused by proper motion is a factor that cannot be ignored when studying their detectability. For NEOs trailing in exposed images, special algorithms are needed to detect them, which is beyond the scope of this paper.

\begin{deluxetable*}{cCCC}[htb!]
\tablecaption{Comparison of different eFoV models\label{tab:eFoV}}
\tablewidth{0pt}
\tablehead{\colhead{Scheme} & \colhead{\textit{Baseline}} & \colhead{\textit{Twinbase}} & \colhead{\textit{Twinaway}}}
\startdata
In tracklets & 1.417 \pm 0.074 & 1.016 \pm 0.008 & 1.014 \pm 0.013\\
Found & 1.128 \pm 0.039 & 1.103 \pm 0.031 & 1.146 \pm 0.034
\enddata
\tablecomments{We calculate the ratio of NEOs numbers of the random eFoV model and the compact eFoV model in 10 runs, and each item in the table represents the average of these 10 ratios and the corresponding $1 \sigma$ error estimate.}
\end{deluxetable*}

Contrary to our analysis in Section \ref{detection}, the random eFoV model will overestimate the search capability of known and unknown NEOs. This is because the tiles in the tile group are closely arranged according to the size of the CCD array, and the FoV circles corresponding to adjacent tiles will overlap. In the random eFoV model, even if a tile group is only observed once that night, many asteroids located in the overlapping area of FoV circle may mistakenly form tracklets. This can be proved from Table \ref{tab:eFoV}. The ratio of the number of NEOs with tracklets in the random eFoV model and the compact eFoV model in the baseline is 1.417, which is much higher than in the other two schemes (1.016 and 1.014). For the random eFoV model, we did find that many simulated objects were detected by different tiles in the same tile group within just a few minutes in the simulated observation data. This shows that our compact eFoV model is more reasonable than the random eFoV model.

\begin{figure*}[htb!]
   \gridline{
        \fig{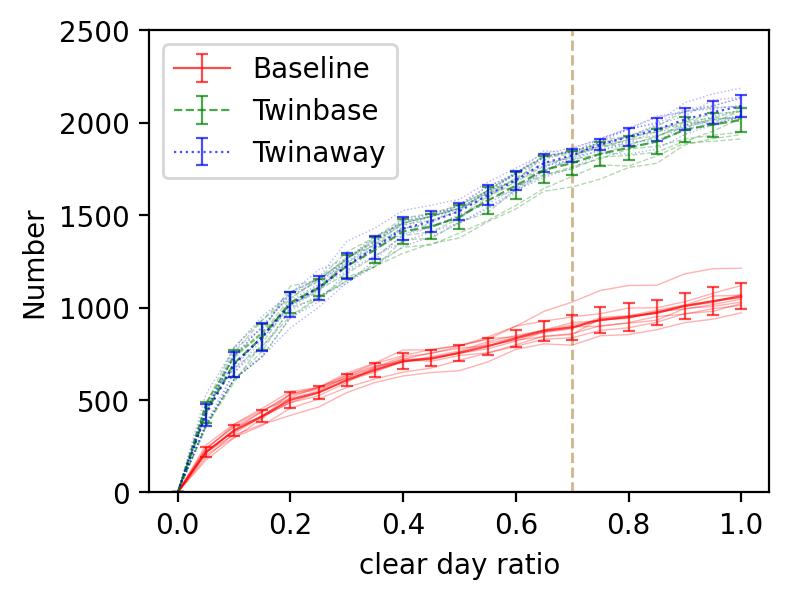}{.45\textwidth}{(a) NEOs with tracklets}
        \fig{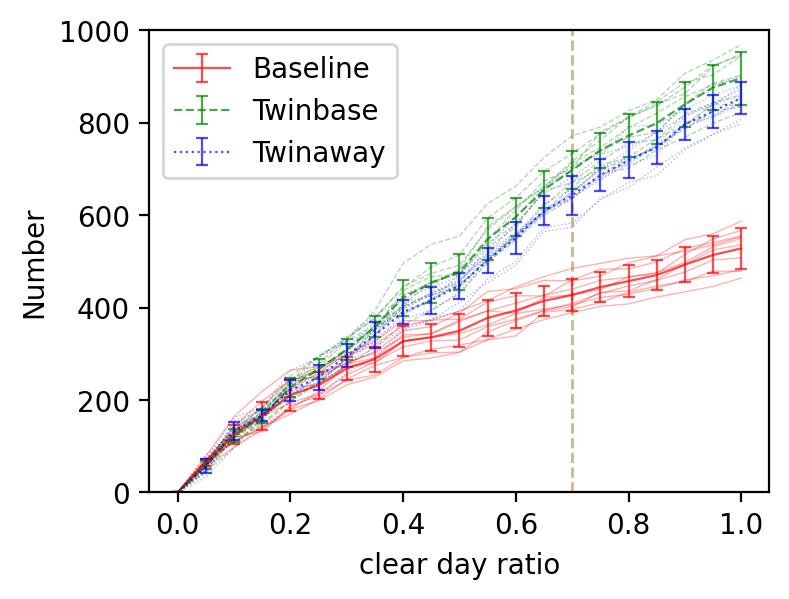}{.45\textwidth}{(b) NEOs found}
    }
    \caption{(a) Number of NEOs with tracklets (not the total number of tracklets), and (b) Number of NEOs found under different clear-day ratios for different simulation schemes in a year. Red (solid), green (dashed) and blue (dotted) represent three schemes: \textit{Baseline}, \textit{Twinbase} and \textit{Twinaway} respectively. Thin solid lines represent the results under 10 different initial conditions, and thick lines give their average values and error bars. Of the nights at the site, 70 \% have clear, photometric conditions.\citep{deng_lenghu_2021} The clear-day ratio of 0.7 is displayed in two panels with vertical dashed lines.}
    \label{fig:clear_day_ratio}
\end{figure*}

For each night, we think the weather is always clear or not clear all night. We randomly assign whether each night is clear or not according to the clear-day ratio and then we got results under different clear-day ratios. As shown in Figure \ref{fig:clear_day_ratio}, we can see that when the clear-day ratio is greater than 0.5, the number of NEOs increases approximately linearly with the clear-day ratio. The concept of a completely clear or not clear night is idealized. When clouds are thin or clouds do not obstruct the observed region, usable data can still be obtained. Therefore, our current approach provides a conservative estimate.

As shown in Table \ref{tab:neo_type}, in the case of a clear-day ratio of 0.7, the average number of NEOs having tracklets is 893.6, 1782.4 and 1824.5 for \textit{Baseline}, \textit{Twinbase} and \textit{Twinaway} respectively, while the average number of NEOs that can be found in a year by the three schemes is 427.3, 697.8 and 642.6, respectively. In comparison with \textit{Baseline}, the search capability of the latter two schemes for known and unknown asteroids is improved by approximately 100\% and 50\%, respectively. Not significantly, \textit{Twinbase} seems to be more conducive to the discovery of unknown NEOs than \textit{Twinaway}, while \textit{Twinaway} seems to have a slight advantage over \textit{Twinbase} in monitoring known NEOs.

\begin{deluxetable*}{ccCCCCC}[htb!]
\tablecaption{Numbers of NEOs with different orbital types identified in mock observations\label{tab:neo_type}}
\tablewidth{0pt}
\tablehead{\colhead{} & \colhead{Scheme} & \colhead{Amors} & \colhead{Apollos} & \colhead{Atens} & \colhead{Total} & \colhead{PHAs}}
\startdata
{} & \textit{Baseline} & 424.1 \pm 26.8 & 427.8 \pm 44.6 & 41.6 \pm 5.6 & 893.6 \pm 68.4 & 132.4 \pm 17.3 \\ 
{In tracklets} & \textit{Twinbase} & 769.1 \pm 34.4 & 902.4 \pm 52.8 & 110.5 \pm 10.2 & 1782.4 \pm 66.3 & 264.6 \pm 23.7 \\ 
{} & \textit{Twinaway} & 775.0 \pm 20.7 & 929.0 \pm 28.1 & 119.7 \pm 13.1 & 1824.5 \pm 35.6 & 278.9 \pm 18.5 \\ 
\hline
{} & \textit{Baseline} & 212.9 \pm 11.8 & 197.3 \pm 26.8 & 17.1 \pm 4.3 & 427.3 \pm 34.0 & 72.6 \pm 11.5 \\ 
{Found} & \textit{Twinbase} & 332.2 \pm 16.5 & 334.7 \pm 31.3 & 30.9 \pm 3.2 & 697.8 \pm 41.5 & 119.7 \pm 15.8 \\ 
{} & \textit{Twinaway} & 303.2 \pm 19.4 & 311.1 \pm 22.5 & 28.1 \pm 5.0 & 642.6 \pm 42.1 & 110.0 \pm 13.6 
\enddata
\tablecomments{Assuming that the clear-day ratio is 0.7, the average numbers and $1 \sigma$ errors of NEOs discoveries in a year of various orbit types under different schemes with known orbits (In tracklets) and unknown orbits (Found). The mean and error of each term are given according to the results of 10 simulations. The average numbers of Atiras never exceed 1.0 and are not listed.}
\end{deluxetable*}

In Granvik's NEOs model \citep{granvik_debiased_2018}, Amors, Apollos, Atens, and Atiras account for 40.4\%, 54.8\%, 3.4\% and 1.4\% respectively, and PHAs account for 0.6\%. As we can see, PHAs are easier to find, because their $H\leq22$ and they are often closer to the earth, making them brighter and easier to find. In addition, there are also observation biases for different orbit types in different schemes and situations. In particular, for the Atiras type, they can hardly be found in WFST's regular survey because of the requirement of the solar altitude angle. Hence, a special twilight survey is needed for Atiras, which is beyond the scope of this paper.

We notice that the DHS tile groups in Figure \ref{fig:cadence} sometimes have four or more visits in one night, because there may be only one DHS tile group to choose from in some cases due to the limitation of the observation conditions. This is not favorable for increasing the number of NEO discoveries and it is usually avoided when making observation plans. We suggest adding some candidate DHS tile groups to make use of this redundant observation time.

\section{Summary}\label{summary}

In this work, the effective field of view, the limiting magnitude, the survey strategy, and the NEO population model are considered to evaluate the search capability of WFST for NEOs with $17<H<25$ through mock observations.

Based on the premise of not violating the WFST regular survey strategy, we present two new schemes to optimize survey scheduling by \citet{chen_basic_2023}. The results of the two new schemes are close, and the search capability of WFST for known and unknown NEOs is improved by 100\% and 50\%, respectively, compared to the original scheme.

We noticed that for NEOs, especially small-sized NEOs, the trailing loss would have a great impact on the results. In addition, we propose a more accurate ``compact" effective field of view model and find that the ``random" effective field of view model will mistakenly overestimate the number of unknown asteroids found in our simulation, resulting in about 10\% overestimation. 

Adopting the clear-day ratio of 0.7, through one year of WFST's regular survey, we estimate that about 1800 NEOs in the optimized scheme can meet the minimum requirements for MPC submission. Even if we search blindly under the same assumption, we can still find more than 600 NEOs. Finally, we suggest adding some candidate DHS sky areas in the rugular survey of WFST to make full use of the redundant observation time of DHS.

\section*{Acknowledgements}
\begin{acknowledgments}
The authors express their appreciation to Ji-An Jiang, Yan-Peng Chen, and Chao Yang for providing specific data regarding the survey plan and extend their gratitude to Ming-Tao Li for his valuable input on this project. In addition, the authors thank the reviewers for their valuable comments and suggestions, which have greatly improved the quality of this manuscript. This work is supported by National Key Research and Development Program of China (2023YFA1608100) and the Strategic Priority Research Program of Chinese Academy of Sciences, Grant No. XDB 41000000. LF gratefully acknowledges the support of the National Natural Science Foundation of China (NSFC, grant No. 12173037, 12233008), the CAS Project for Young Scientists in Basic Research (No. YSBR-092), the Fundamental Research Funds for the Central Universities (WK3440000006) and Cyrus Chun Ying Tang Foundations.
\end{acknowledgments}

\vspace{5mm}
\facilities{WFST: 2.5 m}

\software{Astropy \citep{astropy_collaboration_astropy_2013,astropy_collaboration_astropy_2018,astropy_collaboration_astropy_2022},
          Numpy \citep{harris_array_2020},
          Pandas \citep{the_pandas_development_team_pandas-devpandas_2024},
          PyEphem \citep{rhodes_pyephem_2011},
          PyOrb \citep{kastinen_pyorb_2024},
          MOID.F \citep{wisniowski_fast_2013}
          }

\bibliography{WFST_NEO}{}

\begin{thebibliography}{}
\expandafter\ifx\csname natexlab\endcsname\relax\def\natexlab#1{#1}\fi
\providecommand{\url}[1]{\href{#1}{#1}}
\providecommand{\dodoi}[1]{doi:~\href{http://doi.org/#1}{\nolinkurl{#1}}}
\providecommand{\doeprint}[1]{\href{http://ascl.net/#1}{\nolinkurl{http://ascl.net/#1}}}
\providecommand{\doarXiv}[1]{\href{https://arxiv.org/abs/#1}{\nolinkurl{https://arxiv.org/abs/#1}}}

\bibitem[{Alvarez {et~al.}(1980)Alvarez, Alvarez, Asaro, \&
  Michel}]{alvarez_extraterrestrial_1980}
Alvarez, L.~W., Alvarez, W., Asaro, F., \& Michel, H.~V. 1980, Science, 208,
  1095, \dodoi{10.1126/science.208.4448.1095}

\bibitem[{{Astropy Collaboration} {et~al.}(2013){Astropy Collaboration},
  Robitaille, Tollerud, Greenfield, Droettboom, Bray, Aldcroft, Davis,
  Ginsburg, Price-Whelan, Kerzendorf, Conley, Crighton, Barbary, Muna,
  Ferguson, Grollier, Parikh, Nair, Unther, Deil, Woillez, Conseil, Kramer,
  Turner, Singer, Fox, Weaver, Zabalza, Edwards, Azalee~Bostroem, Burke, Casey,
  Crawford, Dencheva, Ely, Jenness, Labrie, Lim, Pierfederici, Pontzen, Ptak,
  Refsdal, Servillat, \& Streicher}]{astropy_collaboration_astropy_2013}
{Astropy Collaboration}, Robitaille, T.~P., Tollerud, E.~J., {et~al.} 2013,
  Astronomy and Astrophysics, 558, A33, \dodoi{10.1051/0004-6361/201322068}

\bibitem[{{Astropy Collaboration} {et~al.}(2018){Astropy Collaboration},
  Price-Whelan, Sip{\H o}cz, G{\"u}nther, Lim, Crawford, Conseil, Shupe, Craig,
  Dencheva, Ginsburg, VanderPlas, Bradley, P{\'e}rez-Su{\'a}rez, de~Val-Borro,
  Aldcroft, Cruz, Robitaille, Tollerud, Ardelean, Babej, Bach, Bachetti,
  Bakanov, Bamford, Barentsen, Barmby, Baumbach, Berry, Biscani, Boquien,
  Bostroem, Bouma, Brammer, Bray, Breytenbach, Buddelmeijer, Burke, Calderone,
  Cano~Rodr{\'\i}guez, Cara, Cardoso, Cheedella, Copin, Corrales, Crichton,
  D'Avella, Deil, Depagne, Dietrich, Donath, Droettboom, Earl, Erben, Fabbro,
  Ferreira, Finethy, Fox, Garrison, Gibbons, Goldstein, Gommers, Greco,
  Greenfield, Groener, Grollier, Hagen, Hirst, Homeier, Horton, Hosseinzadeh,
  Hu, Hunkeler, Ivezi{\'c}, Jain, Jenness, Kanarek, Kendrew, Kern, Kerzendorf,
  Khvalko, King, Kirkby, Kulkarni, Kumar, Lee, Lenz, Littlefair, Ma, Macleod,
  Mastropietro, McCully, Montagnac, Morris, Mueller, Mumford, Muna, Murphy,
  Nelson, Nguyen, Ninan, N{\"o}the, Ogaz, Oh, Parejko, Parley, Pascual, Patil,
  Patil, Plunkett, Prochaska, Rastogi, Reddy~Janga, Sabater, Sakurikar,
  Seifert, Sherbert, Sherwood-Taylor, Shih, Sick, Silbiger, Singanamalla,
  Singer, Sladen, Sooley, Sornarajah, Streicher, Teuben, Thomas, Tremblay,
  Turner, Terr{\'o}n, van Kerkwijk, de~la Vega, Watkins, Weaver, Whitmore,
  Woillez, Zabalza, \& {Astropy
  Contributors}}]{astropy_collaboration_astropy_2018}
{Astropy Collaboration}, Price-Whelan, A.~M., Sip{\H o}cz, B.~M., {et~al.}
  2018, The Astronomical Journal, 156, 123, \dodoi{10.3847/1538-3881/aabc4f}

\bibitem[{{Astropy Collaboration} {et~al.}(2022){Astropy Collaboration},
  Price-Whelan, Lim, Earl, Starkman, Bradley, Shupe, Patil, Corrales, Brasseur,
  N{\"o}the, Donath, Tollerud, Morris, Ginsburg, Vaher, Weaver, Tocknell,
  Jamieson, van Kerkwijk, Robitaille, Merry, Bachetti, G{\"u}nther, Aldcroft,
  Alvarado-Montes, Archibald, B{\'o}di, Bapat, Barentsen, Baz{\'a}n, Biswas,
  Boquien, Burke, Cara, Cara, Conroy, Conseil, Craig, Cross, Cruz, D'Eugenio,
  Dencheva, Devillepoix, Dietrich, Eigenbrot, Erben, Ferreira, Foreman-Mackey,
  Fox, Freij, Garg, Geda, Glattly, Gondhalekar, Gordon, Grant, Greenfield,
  Groener, Guest, Gurovich, Handberg, Hart, Hatfield-Dodds, Homeier,
  Hosseinzadeh, Jenness, Jones, Joseph, Kalmbach, Karamehmetoglu,
  Ka{\l}uszy{\'n}ski, Kelley, Kern, Kerzendorf, Koch, Kulumani, Lee, Ly, Ma,
  MacBride, Maljaars, Muna, Murphy, Norman, O'Steen, Oman, Pacifici, Pascual,
  Pascual-Granado, Patil, Perren, Pickering, Rastogi, Roulston, Ryan, Rykoff,
  Sabater, Sakurikar, Salgado, Sanghi, Saunders, Savchenko, Schwardt,
  Seifert-Eckert, Shih, Jain, Shukla, Sick, Simpson, Singanamalla, Singer,
  Singhal, Sinha, Sip{\H o}cz, Spitler, Stansby, Streicher, {\v S}umak,
  Swinbank, Taranu, Tewary, Tremblay, de~Val-Borro, Van~Kooten, Vasovi{\'c},
  Verma, de~Miranda~Cardoso, Williams, Wilson, Winkel, Wood-Vasey, Xue,
  Yoachim, Zhang, Zonca, \& {Astropy Project
  Contributors}}]{astropy_collaboration_astropy_2022}
{Astropy Collaboration}, Price-Whelan, A.~M., Lim, P.~L., {et~al.} 2022, The
  Astrophysical Journal, 935, 167, \dodoi{10.3847/1538-4357/ac7c74}

\bibitem[{Bellm {et~al.}(2019)Bellm, Kulkarni, Graham, Dekany, Smith, Riddle,
  Masci, Helou, Prince, Adams, Barbarino, Barlow, Bauer, Beck, Belicki, Biswas,
  Blagorodnova, Bodewits, Bolin, Brinnel, Brooke, Bue, Bulla, Burruss, Cenko,
  Chang, Connolly, Coughlin, Cromer, Cunningham, De, Delacroix, Desai, Duev,
  Eadie, Farnham, Feeney, Feindt, Flynn, Franckowiak, Frederick, Fremling,
  Gal-Yam, Gezari, Giomi, Goldstein, Golkhou, Goobar, Groom, Hacopians, Hale,
  Henning, Ho, Hover, Howell, Hung, Huppenkothen, Imel, Ip, Ivezi{\'c},
  Jackson, Jones, Juric, Kasliwal, Kaspi, Kaye, Kelley, Kowalski, Kramer,
  Kupfer, Landry, Laher, Lee, Lin, Lin, Lunnan, Giomi, Mahabal, Mao, Miller,
  Monkewitz, Murphy, Ngeow, Nordin, Nugent, Ofek, Patterson, Penprase, Porter,
  Rauch, Rebbapragada, Reiley, Rigault, Rodriguez, Roestel, Rusholme, Santen,
  Schulze, Shupe, Singer, Soumagnac, Stein, Surace, Sollerman, Szkody, Taddia,
  Terek, Van~Sistine, Van~Velzen, Vestrand, Walters, Ward, Ye, Yu, Yan, \&
  Zolkower}]{bellm_zwicky_2019}
Bellm, E.~C., Kulkarni, S.~R., Graham, M.~J., {et~al.} 2019, PASP, 131, 018002,
  \dodoi{10.1088/1538-3873/aaecbe}

\bibitem[{Bottke(2002)}]{bottke_debiased_2002}
Bottke, W. 2002, Icarus, 156, 399, \dodoi{10.1006/icar.2001.6788}

\bibitem[{Bottke {et~al.}(2000)Bottke, Jedicke, Morbidelli, Petit, \&
  Gladman}]{bottke_understanding_2000}
Bottke, W.~F., Jedicke, R., Morbidelli, A., Petit, J.-M., \& Gladman, B. 2000,
  Science, 288, 2190, \dodoi{10.1126/science.288.5474.2190}

\bibitem[{Bowell {et~al.}(1989)Bowell, Hapke, Domingue, Lumme, Peltoniemi, \&
  Harris}]{bowell_application_1989}
Bowell, E., Hapke, B., Domingue, D., {et~al.} 1989, Application of photometric
  models to asteroids. (Tucson, AZ: University of Arizona Press).
\newblock \url{https://ui.adsabs.harvard.edu/abs/1989aste.conf..524B}

\bibitem[{Brown {et~al.}(2013)Brown, Assink, Astiz, Blaauw, Boslough, Borovi{\v
  c}ka, Brachet, Brown, Campbell-Brown, Ceranna, Cooke, De~Groot-Hedlin, Drob,
  Edwards, Evers, Garces, Gill, Hedlin, Kingery, Laske, Le~Pichon, Mialle,
  Moser, Saffer, Silber, Smets, Spalding, Spurn{\'y}, Tagliaferri, Uren, Weryk,
  Whitaker, \& Krzeminski}]{brown_500-kiloton_2013}
Brown, P.~G., Assink, J.~D., Astiz, L., {et~al.} 2013, Nature, 503, 238,
  \dodoi{10.1038/nature12741}

\bibitem[{Cai {et~al.}(2025)Cai, Xu, Fan, Wan, Kong, Hu, Jiang, Hu, Zhu, Li,
  Lin, Fang, Xue, Zhen, \& Wang}]{cai_25-meter_2025}
Cai, M., Xu, Z., Fan, L., {et~al.} 2025, The 2.5-meter {Wide} {Field} {Survey}
  {Telescope} {Real}-time {Data} {Processing} {Pipeline} {I}: {From} raw data
  to alert distribution,  arXiv, \dodoi{10.48550/arXiv.2501.15018}

\bibitem[{Chapman(2004)}]{chapman_hazard_2004}
Chapman, C.~R. 2004, Earth and Planetary Science Letters, 222, 1,
  \dodoi{10.1016/j.epsl.2004.03.004}

\bibitem[{Chen {et~al.}(2023)Chen, Jiang, Luo, Zheng, Fang, Yang, Hong, \&
  L{\"u}}]{chen_basic_2023}
Chen, Y.-P., Jiang, J.-A., Luo, W.-T., {et~al.} 2023, Res. Astron. Astrophys.,
  24, 015003, \dodoi{10.1088/1674-4527/ad07cd}

\bibitem[{Christensen {et~al.}(2019)Christensen, Africano, Farneth, Fuls,
  Gibbs, Grauer, Groeller, Kowalski, Larson, Leonard, Prune, Rankin, Seaman, \&
  Shelly}]{christensen_catalina_2019}
Christensen, E., Africano, B., Farneth, G., {et~al.} 2019, in {EPSC}-{DPS}
  {Joint} {Meeting} 2019, Vol. 2019, EPSC--DPS2019--1912.
\newblock \url{https://ui.adsabs.harvard.edu/abs/2019EPSC...13.1912C}

\bibitem[{Deienno {et~al.}(2025)Deienno, Denneau, Nesvorn{\'y},
  Vokrouhlick{\'y}, Bottke, Jedicke, Naidu, Chesley, Farnocchia, \&
  Chodas}]{deienno_debiased_2025}
Deienno, R., Denneau, L., Nesvorn{\'y}, D., {et~al.} 2025, Icarus, 425, 116316,
  \dodoi{10.1016/j.icarus.2024.116316}

\bibitem[{Deng {et~al.}(2021)Deng, Yang, Chen, He, Liu, Zhang, Zhang, Wang,
  Liu, Ren, Luo, Yan, Tian, \& Pan}]{deng_lenghu_2021}
Deng, L., Yang, F., Chen, X., {et~al.} 2021, Nature, 596, 353,
  \dodoi{10.1038/s41586-021-03711-z}

\bibitem[{Denneau {et~al.}(2013)Denneau, Jedicke, Grav, Granvik, Kubica,
  Milani, Vere{\v s}, Wainscoat, Chang, Pierfederici, Kaiser, Chambers,
  Heasley, Magnier, Price, Myers, Kleyna, Hsieh, Farnocchia, Waters, Sweeney,
  Green, Bolin, Burgett, Morgan, Tonry, Hodapp, Chastel, Chesley, Fitzsimmons,
  Holman, Spahr, Tholen, Williams, Abe, Armstrong, Bressi, Holmes, Lister,
  McMillan, Micheli, Ryan, Ryan, \& Scotti}]{denneau_pan-starrs_2013}
Denneau, L., Jedicke, R., Grav, T., {et~al.} 2013, Publications of the
  Astronomical Society of the Pacific, 125, 357, \dodoi{10.1086/670337}

\bibitem[{Fasbender \& Nidever(2021)}]{fasbender_exploring_2021}
Fasbender, K.~M., \& Nidever, D.~L. 2021, AJ, 162, 244,
  \dodoi{10.3847/1538-3881/ac2230}

\bibitem[{Granvik {et~al.}(2016)Granvik, Morbidelli, Jedicke, Bolin, Bottke,
  Beshore, Vokrouhlick{\'y}, Delb{\`o}, \&
  Michel}]{granvik_super-catastrophic_2016}
Granvik, M., Morbidelli, A., Jedicke, R., {et~al.} 2016, Nature, 530, 303,
  \dodoi{10.1038/nature16934}

\bibitem[{Granvik {et~al.}(2018)Granvik, Morbidelli, Jedicke, Bolin, Bottke,
  Beshore, Vokrouhlick{\'y}, Nesvorn{\'y}, \& Michel}]{granvik_debiased_2018}
---. 2018, Icarus, 312, 181, \dodoi{10.1016/j.icarus.2018.04.018}

\bibitem[{Harris {et~al.}(2020)Harris, Millman, Van Der~Walt, Gommers,
  Virtanen, Cournapeau, Wieser, Taylor, Berg, Smith, Kern, Picus, Hoyer,
  Van~Kerkwijk, Brett, Haldane, Del~R{\'\i}o, Wiebe, Peterson,
  G{\'e}rard-Marchant, Sheppard, Reddy, Weckesser, Abbasi, Gohlke, \&
  Oliphant}]{harris_array_2020}
Harris, C.~R., Millman, K.~J., Van Der~Walt, S.~J., {et~al.} 2020, Nature, 585,
  357, \dodoi{10.1038/s41586-020-2649-2}

\bibitem[{Heinze {et~al.}(2021)Heinze, Denneau, Tonry, Smartt, Erasmus,
  Fitzsimmons, Robinson, Weiland, Flewelling, Stalder, Rest, \&
  Young}]{heinze_neo_2021}
Heinze, A.~N., Denneau, L., Tonry, J.~L., {et~al.} 2021, Planet. Sci. J., 2,
  12, \dodoi{10.3847/PSJ/abd325}

\bibitem[{Holman {et~al.}(2018)Holman, Payne, Blankley, Janssen, \&
  Kuindersma}]{holman_heliolinc_2018}
Holman, M.~J., Payne, M.~J., Blankley, P., Janssen, R., \& Kuindersma, S. 2018,
  AJ, 156, 135, \dodoi{10.3847/1538-3881/aad69a}

\bibitem[{Hu {et~al.}(2022)Hu, Hu, Jiang, Xiao, Fan, Wei, \&
  Wu}]{hu_prospects_2022}
Hu, M., Hu, L., Jiang, J.-a., {et~al.} 2022, Universe, 9, 7,
  \dodoi{10.3390/universe9010007}

\bibitem[{JeongAhn \& Malhotra(2014)}]{jeongahn_non-uniform_2014}
JeongAhn, Y., \& Malhotra, R. 2014, Icarus, 229, 236,
  \dodoi{10.1016/j.icarus.2013.10.030}

\bibitem[{Jones {et~al.}(2018)Jones, Slater, Moeyens, Allen, Axelrod, Cook,
  Ivezi{\'c}, Juri{\'c}, Myers, \& Petry}]{jones_large_2018}
Jones, R.~L., Slater, C.~T., Moeyens, J., {et~al.} 2018, Icarus, 303, 181,
  \dodoi{10.1016/j.icarus.2017.11.033}

\bibitem[{Kastinen(2024)}]{kastinen_pyorb_2024}
Kastinen, D. 2024, {PyOrb} --- {Keplerian} orbit functions in {Python}.
\newblock \url{https://danielk.developer.irf.se/pyorb/}

\bibitem[{Krisciunas \& Schaefer(1991)}]{krisciunas_model_1991}
Krisciunas, K., \& Schaefer, B.~E. 1991, Publications of the Astronomical
  Society of the Pacific, 103, 1033, \dodoi{10.1086/132921}

\bibitem[{Lei {et~al.}(2023)Lei, Zhu, Kong, Wang, Zheng, Shi, Fan, \&
  Liu}]{lei_limiting_2023}
Lei, L., Zhu, Q.-F., Kong, X., {et~al.} 2023, Res. Astron. Astrophys., 23,
  035013, \dodoi{10.1088/1674-4527/acb877}

\bibitem[{Liang {et~al.}(2023)Liang, Liu, Lei, \&
  Zhao}]{liang_kilonova-targeting_2023}
Liang, R., Liu, Z., Lei, L., \& Zhao, W. 2023, Universe, 10, 10,
  \dodoi{10.3390/universe10010010}

\bibitem[{Lin {et~al.}(2022)Lin, Jiang, \& Kong}]{lin_prospects_2022}
Lin, Z., Jiang, N., \& Kong, X. 2022, Monthly Notices of the Royal Astronomical
  Society, 513, 2422, \dodoi{10.1093/mnras/stac946}

\bibitem[{Liu {et~al.}(2025)Liu, Fan, Hu, Lu, Lu, Xu, Zhu, Wang, \&
  Kong}]{liu_classification_2025}
Liu, Y., Fan, L., Hu, L., {et~al.} 2025, A\&A, 693, A105,
  \dodoi{10.1051/0004-6361/202348581}

\bibitem[{Liu {et~al.}(2023)Liu, Lin, Yu, Wang, Mourani, Zhao, \&
  Dai}]{liu_target--opportunity_2023}
Liu, Z.-Y., Lin, Z.-Y., Yu, J.-M., {et~al.} 2023, ApJ, 947, 59,
  \dodoi{10.3847/1538-4357/acc73b}

\bibitem[{Mainzer {et~al.}(2014)Mainzer, Bauer, Grav, Masiero, Cutri, Wright,
  Nugent, Stevenson, Clyne, Cukrov, \& Masci}]{mainzer_population_2014}
Mainzer, A., Bauer, J., Grav, T., {et~al.} 2014, The Astrophysical Journal,
  784, 110, \dodoi{10.1088/0004-637X/784/2/110}

\bibitem[{Mainzer {et~al.}(2023)Mainzer, Masiero, Abell, Bauer, Bottke,
  Buratti, Carey, Cotto-Figueroa, Cutri, Dahlen, Eisenhardt, Fernandez,
  Furfaro, Grav, Hoffman, Kelley, Kim, Kirkpatrick, Lawler, Lilly, Liu,
  Marocco, Marsh, Masci, McMurtry, Pourrahmani, Reinhart, Ressler, Satpathy,
  Schambeau, Sonnett, Spahr, Surace, Vaquero, Wright, Zengilowski, \& {NEO
  Surveyor Mission Team}}]{mainzer_near-earth_2023}
Mainzer, A.~K., Masiero, J.~R., Abell, P.~A., {et~al.} 2023, Planet. Sci. J.,
  4, 224, \dodoi{10.3847/PSJ/ad0468}

\bibitem[{Masci {et~al.}(2019)Masci, Laher, Rusholme, Shupe, Groom, Surace,
  Jackson, Monkewitz, Beck, Flynn, Terek, Landry, Hacopians, Desai, Howell,
  Brooke, Imel, Wachter, Ye, Lin, Cenko, Cunningham, Rebbapragada, Bue, Miller,
  Mahabal, Bellm, Patterson, Juri{\'c}, Golkhou, Ofek, Walters, Graham,
  Kasliwal, Dekany, Kupfer, Burdge, Cannella, Barlow, Sistine, Giomi, Fremling,
  Blagorodnova, Levitan, Riddle, Smith, Helou, Prince, \&
  Kulkarni}]{masci_zwicky_2019}
Masci, F.~J., Laher, R.~R., Rusholme, B., {et~al.} 2019, PASP, 131, 018003,
  \dodoi{10.1088/1538-3873/aae8ac}

\bibitem[{Masiero {et~al.}(2021)Masiero, Mainzer, Bauer, Cutri, Grav,
  Pittichov{\'a}, Kramer, \& Wright}]{masiero_characterization_2021}
Masiero, J., Mainzer, A., Bauer, J.~M., {et~al.} 2021, in 7th {IAA} {Planetary}
  {Defense} {Conference}, 34.
\newblock \url{https://ui.adsabs.harvard.edu/abs/2021plde.confE..34M}

\bibitem[{Moeyens {et~al.}(2021)Moeyens, Juri{\'c}, Ford, Bekte{\v s}evi{\'c},
  Connolly, Eggl, Ivezi{\'c}, Jones, Kalmbach, \&
  Smotherman}]{moeyens_thor_2021}
Moeyens, J., Juri{\'c}, M., Ford, J., {et~al.} 2021, AJ, 162, 143,
  \dodoi{10.3847/1538-3881/ac042b}

\bibitem[{Montagner {et~al.}(2023)Montagner, Peloton, Carry, Desmars,
  Hestroffer, Mendez, Perlbarg, \& Thuillot}]{montagner_enabling_2023}
Montagner, R.~L., Peloton, J., Carry, B., {et~al.} 2023, A\&A, 680, A17,
  \dodoi{10.1051/0004-6361/202346905}

\bibitem[{Nesvorn{\'y} {et~al.}(2023)Nesvorn{\'y}, Deienno, Bottke, Jedicke,
  Naidu, Chesley, Chodas, Granvik, Vokrouhlick{\'y}, Bro{\v z}, Morbidelli,
  Christensen, Shelly, \& Bolin}]{nesvorny_neomod_2023}
Nesvorn{\'y}, D., Deienno, R., Bottke, W.~F., {et~al.} 2023, AJ, 166, 55,
  \dodoi{10.3847/1538-3881/ace040}

\bibitem[{Nesvorn{\'y} {et~al.}(2024{\natexlab{a}})Nesvorn{\'y},
  Vokrouhlick{\'y}, Shelly, Deienno, Bottke, Christensen, Jedicke, Naidu,
  Chesley, Chodas, Farnocchia, \& Granvik}]{nesvorny_neomod_2024}
Nesvorn{\'y}, D., Vokrouhlick{\'y}, D., Shelly, F., {et~al.}
  2024{\natexlab{a}}, Icarus, 411, 115922, \dodoi{10.1016/j.icarus.2023.115922}

\bibitem[{Nesvorn{\'y} {et~al.}(2024{\natexlab{b}})Nesvorn{\'y},
  Vokrouhlick{\'y}, Shelly, Deienno, Bottke, Fuls, Jedicke, Naidu, Chesley,
  Chodas, Farnocchia, \& Delbo}]{nesvorny_neomod_2024-1}
---. 2024{\natexlab{b}}, Icarus, 417, 116110,
  \dodoi{10.1016/j.icarus.2024.116110}

\bibitem[{Nicholson {et~al.}(2024)Nicholson, Powell, Gulick, Kenkmann, Bray,
  Duarte, \& Collins}]{nicholson_3d_2024}
Nicholson, U., Powell, W., Gulick, S., {et~al.} 2024, Commun Earth Environ, 5,
  547, \dodoi{10.1038/s43247-024-01700-4}

\bibitem[{Noll {et~al.}(2012)Noll, Kausch, Barden, Jones, Szyszka, Kimeswenger,
  \& Vinther}]{noll_atmospheric_2012}
Noll, S., Kausch, W., Barden, M., {et~al.} 2012, A\&A, 543, A92,
  \dodoi{10.1051/0004-6361/201219040}

\bibitem[{Popova {et~al.}(2013)Popova, Jenniskens, Emel'yanenko, Kartashova,
  Biryukov, Khaibrakhmanov, Shuvalov, Rybnov, Dudorov, Grokhovsky, Badyukov,
  Yin, Gural, Albers, Granvik, Evers, Kuiper, Kharlamov, Solovyov, Rusakov,
  Korotkiy, Serdyuk, Korochantsev, Larionov, Glazachev, Mayer, Gisler,
  Gladkovsky, Wimpenny, Sanborn, Yamakawa, Verosub, Rowland, Roeske, Botto,
  Friedrich, Zolensky, Le, Ross, Ziegler, Nakamura, Ahn, Lee, Zhou, Li, Li,
  Liu, Tang, Hiroi, Sears, Weinstein, Vokhmintsev, Ishchenko, Schmitt-Kopplin,
  Hertkorn, Nagao, Haba, Komatsu, Mikouchi, \& {(the Chelyabinsk Airburst
  Consortium)}}]{popova_chelyabinsk_2013}
Popova, O.~P., Jenniskens, P., Emel'yanenko, V., {et~al.} 2013, Science, 342,
  1069, \dodoi{10.1126/science.1242642}

\bibitem[{Rhodes(2011)}]{rhodes_pyephem_2011}
Rhodes, B.~C. 2011, Astrophysics Source Code Library, ascl:1112.014.
\newblock \url{https://ui.adsabs.harvard.edu/abs/2011ascl.soft12014R}

\bibitem[{Rozenberg(1966)}]{rozenberg_twilight_1966}
Rozenberg, G.~V. 1966, Twilight (Boston, MA: Springer US),
  \dodoi{10.1007/978-1-4899-6353-6}

\bibitem[{Schlegel {et~al.}(1998)Schlegel, Finkbeiner, \&
  Davis}]{schlegel_maps_1998}
Schlegel, D.~J., Finkbeiner, D.~P., \& Davis, M. 1998, The Astrophysical
  Journal, 500, 525, \dodoi{10.1086/305772}

\bibitem[{Skillen(2002)}]{skillen_new_2002}
Skillen, I. 2002, The Newsletter of the Isaac Newton Group of Telescopes, 6,
  38.
\newblock \url{https://ui.adsabs.harvard.edu/abs/2002INGN....6...38S}

\bibitem[{Stuart \& Binzel(2004)}]{stuart_bias-corrected_2004}
Stuart, J.~S., \& Binzel, R.~P. 2004, Icarus, 170, 295,
  \dodoi{10.1016/j.icarus.2004.03.018}

\bibitem[{{The pandas development
  team}(2024)}]{the_pandas_development_team_pandas-devpandas_2024}
{The pandas development team}. 2024, pandas-dev/pandas: {Pandas},  Zenodo,
  \dodoi{10.5281/ZENODO.13819579}

\bibitem[{Wainscoat(2016)}]{wainscoat_pan-starrs_2016}
Wainscoat, R. 2016, in 2016 {IEEE} {Aerospace} {Conference}, 1--6,
  \dodoi{10.1109/AERO.2016.7500568}

\bibitem[{Wang {et~al.}(2022)Wang, Yu, Liu, Zhao, \& Lu}]{wang_study_2022}
Wang, H.-Y., Yu, J.-M., Liu, Z.-Y., Zhao, W., \& Lu, Y.-J. 2022, SSPMA, 53, 1,
  \dodoi{10.1360/SSPMA-2022-0252}

\bibitem[{Wang {et~al.}(2025)Wang, Fu, Lu, Fan, Cai, Xu, Kong, Zhao, Li, Liu,
  Zhu, Zhou, Wan, Cheng, Jiang, Li, Liang, Liu, Luo, Lou, Wang, Wang, Wang,
  Xue, Zhang, \& Zhao}]{wang_heliocentric-orbiting_2025}
Wang, S.-H., Fu, B.-X., Lu, J.-Q., {et~al.} 2025, A {Heliocentric}-orbiting
  {Objects} {Processing} {System} ({HOPS}) for the {Wide} {Field} {Survey}
  {Telescope}: {Architecture}, {Processing} {Workflow}, and {Preliminary}
  {Results},  arXiv, \dodoi{10.48550/arXiv.2501.17472}

\bibitem[{Wang {et~al.}(2023)Wang, Liu, Cai, Geng, Fang, He, Jiang, Jiang,
  Kong, Li, Li, Luo, Pan, Wu, Yang, Yu, Zheng, Zhu, Cai, Chen, Chen, Dai, Fan,
  Fan, Fang, He, Hu, Hu, Jin, Jiang, Li, Li, Li, Liang, Lin, Liu, Liu, Liu,
  Liu, Liu, Lou, Qu, Sheng, Shi, Shu, Su, Sun, Wang, Wang, Wang, Wang, Wei,
  Wei, Xue, Yan, Yang, Yuan, Yuan, Zhang, Zhang, Zhao, \&
  Zhao}]{wang_sciences_2023}
Wang, T., Liu, G., Cai, Z., {et~al.} 2023, Sci. China Phys. Mech. Astron., 66,
  109512, \dodoi{10.1007/s11433-023-2197-5}

\bibitem[{Wi{\'s}niowski \& Rickman(2013)}]{wisniowski_fast_2013}
Wi{\'s}niowski, T., \& Rickman, H. 2013, Acta Astronomica, 63, 293.
\newblock \url{https://ui.adsabs.harvard.edu/abs/2013AcA....63..293W}

\bibitem[{Xu {et~al.}(2022)Xu, Zhu, Li, Li, Zheng, Qiu, \& Zhao}]{xu_new_2022}
Xu, X.-H., Zhu, Q.-F., Li, X.-Z., {et~al.} 2022, PASP, 134, 114507,
  \dodoi{10.1088/1538-3873/ac9e1b}

\end{thebibliography}
\bibliographystyle{aasjournal}

\end{document}